\documentclass[sigconf,natbib=true,anonymous=false]{acmart}
\AtBeginDocument{%
  \providecommand\BibTeX{{%
    \normalfont B\kern-0.5em{\scshape i\kern-0.25em b}\kern-0.8em\TeX}}}

\usepackage{booktabs} 

\usepackage[english]{babel}
\usepackage{moresize}
\usepackage{amsmath}
\usepackage{balance}
\usepackage{comment}
\usepackage{paralist}
\usepackage{bm}
\usepackage{pgfplots}
\usetikzlibrary{pgfplots.dateplot}

\usepackage{flushend}
\usepackage[english]{babel}
\usepackage[latin1]{inputenc}
\usepackage{mathrsfs}
\usepackage{graphicx}

\usepackage{amssymb}
\usepackage{amsfonts}
\usepackage{url}
\usepackage{longtable}
\usepackage{mathtools}
\usepackage{rotating}
\usepackage{multirow}
\usepackage{mathrsfs}
\usepackage{subcaption}
\usepackage{enumitem}
\usepackage[linesnumbered,algoruled,boxed,lined]{algorithm2e}
\usepackage{hyperref}
\usepackage{pgfplots}
\usetikzlibrary{pgfplots.dateplot}
\usepackage{filecontents}
\usepackage{threeparttable}
\usepackage[skip=0.5\baselineskip]{caption}
\definecolor{tblue}{RGB}{31,119,180}
\definecolor{torange}{RGB}{255,127,14}
\definecolor{tgreen}{RGB}{44,160,44}
\definecolor{tred}{RGB}{214,39,40}
\definecolor{tpurple}{RGB}{148,103,189}

\newcommand{\hide}[1]{} 

\SetKwInput{KwInput}{Input}                
\SetKwInput{KwOutput}{Output}

\setcopyright{none}
 \setcopyright{acmcopyright}
 \copyrightyear{2018}
 \acmYear{2018}
 \acmDOI{XXXXXXX.XXXXXXX}

 \acmConference[Conference acronym 'XX]{Make sure to enter the correct
   conference title from your rights confirmation emai}{June 03--05,
   2018}{Woodstock, NY}

 \acmPrice{15.00}
 \acmISBN{978-1-4503-XXXX-X/18/06}

\begin{document}

\copyrightyear{2023}
\acmYear{2023}
\setcopyright{acmlicensed}\acmConference[CIKM '23]{Proceedings of the 32nd ACM International Conference on Information and Knowledge Management}{October 21--25, 2023}{Birmingham, United Kingdom}
\acmBooktitle{Proceedings of the 32nd ACM International Conference on Information and Knowledge Management (CIKM '23), October 21--25, 2023, Birmingham, United Kingdom}
\acmPrice{15.00}
\acmDOI{10.1145/3583780.3614917}
\acmISBN{979-8-4007-0124-5/23/10}

\title{How Expressive are Graph Neural Networks in Recommendation?}

\author{Xuheng Cai}
\affiliation{%
  \institution{The University of Hong Kong}
  \country{Hong Kong, China}
}
\email{rickcai@connect.hku.hk}

\author{Lianghao Xia}
\affiliation{%
  \institution{The University of Hong Kong}
  \country{Hong Kong, China}
}
\email{aka_xia@foxmail.com}

\author{Xubin Ren}
\affiliation{%
  \institution{The University of Hong Kong}
  \country{Hong Kong, China}
}
\email{xubinrencs@gmail.com}

\author{Chao Huang}
\authornote{Chao Huang is the corresponding author.}
\affiliation{%
  \institution{The University of Hong Kong}
  \country{Hong Kong, China}
}
\email{chaohuang75@gmail.com}

\renewcommand{\shortauthors}{Xuheng Cai, Lianghao Xia, Xubin Ren, \& Chao Huang}

\begin{abstract}

Graph Neural Networks (GNNs) have demonstrated superior performance in various graph learning tasks, including recommendation, where they explore user-item collaborative filtering signals within graphs. However, despite their empirical effectiveness in state-of-the-art recommender models, theoretical formulations of their capability are scarce. Recently, researchers have explored the expressiveness of GNNs, demonstrating that message passing GNNs are at most as powerful as the Weisfeiler-Lehman test, and that GNNs combined with random node initialization are universal. Nevertheless, the concept of "expressiveness" for GNNs remains vaguely defined. Most existing works adopt the graph isomorphism test as the metric of expressiveness, but this graph-level task may not effectively assess a model's ability in recommendation, where the objective is to distinguish nodes of different closeness. In this paper, we provide a comprehensive theoretical analysis of the expressiveness of GNNs in recommendation, considering three levels of expressiveness metrics: graph isomorphism (graph-level), node automorphism (node-level), and topological closeness (link-level). We propose the topological closeness metric to evaluate GNNs' ability to capture the structural distance between nodes, which closely aligns with the recommendation objective. To validate the effectiveness of this new metric in evaluating recommendation performance, we introduce a learning-less GNN algorithm that is optimal on the new metric and can be optimal on the node-level metric with suitable modification. We conduct extensive experiments comparing the proposed algorithm against various types of state-of-the-art GNN models to explore the effectiveness of the new metric in the recommendation task. For the sake of reproducibility, implementation codes are available at \url{https://github.com/HKUDS/GTE}.

\end{abstract}




\maketitle
\section{Introduction}
\label{sec:intro}

Graph Neural Networks (GNNs) are powerful tools for encoding structural information in graphs and have found applications in various real-world domains, including biochemistry \cite{fout2017protein, gilmer2017neural}, knowledge modeling \cite{schlichtkrull2018modeling}, and social networks \cite{hamilton2017representation, xu2018powerful, abboud2020surprising}. One particular application that has gained significant attention is recommender systems \cite{ying2018graph, wang2019neural, he2020lightgcn}, which utilize collaborative filtering signals from users' past interactions to deliver personalized recommendations in online video streaming \cite{guo2020survey}, e-commerce platforms \cite{jiang2019understanding}, and social media \cite{ko2022survey}. GNN-based recommender systems have demonstrated superior performance on real-world datasets, especially with the recent advancements in Graph Contrastive Learning (GCL), which enhance the GNN learning process through self-supervision signals \cite{wu2021self, xia2022hypergraph, yu2022graph, ren2023disentangled}. However, while existing works validate the effectiveness of GNN-based recommenders through empirical performance, the theoretical understanding of GNNs' capabilities in addressing recommendation tasks remains limited \cite{ferrari2019we, ferrari2020critically, ferrari2020methodological, ferrari2021troubling}. Among the existing theoretical works, Shen et al. \cite{shen2021powerful} analyze graph convolution from the perspective of graph signal processing, providing insights into how GNN-based recommenders extract collaborative signals from graphs. Nevertheless, there is still a lack of comprehensive guarantees on the potential of GNNs for graph-related tasks.

\begin{figure}
        \begin{subfigure}{\linewidth}
        \centering
            \includegraphics[width=0.85\textwidth]{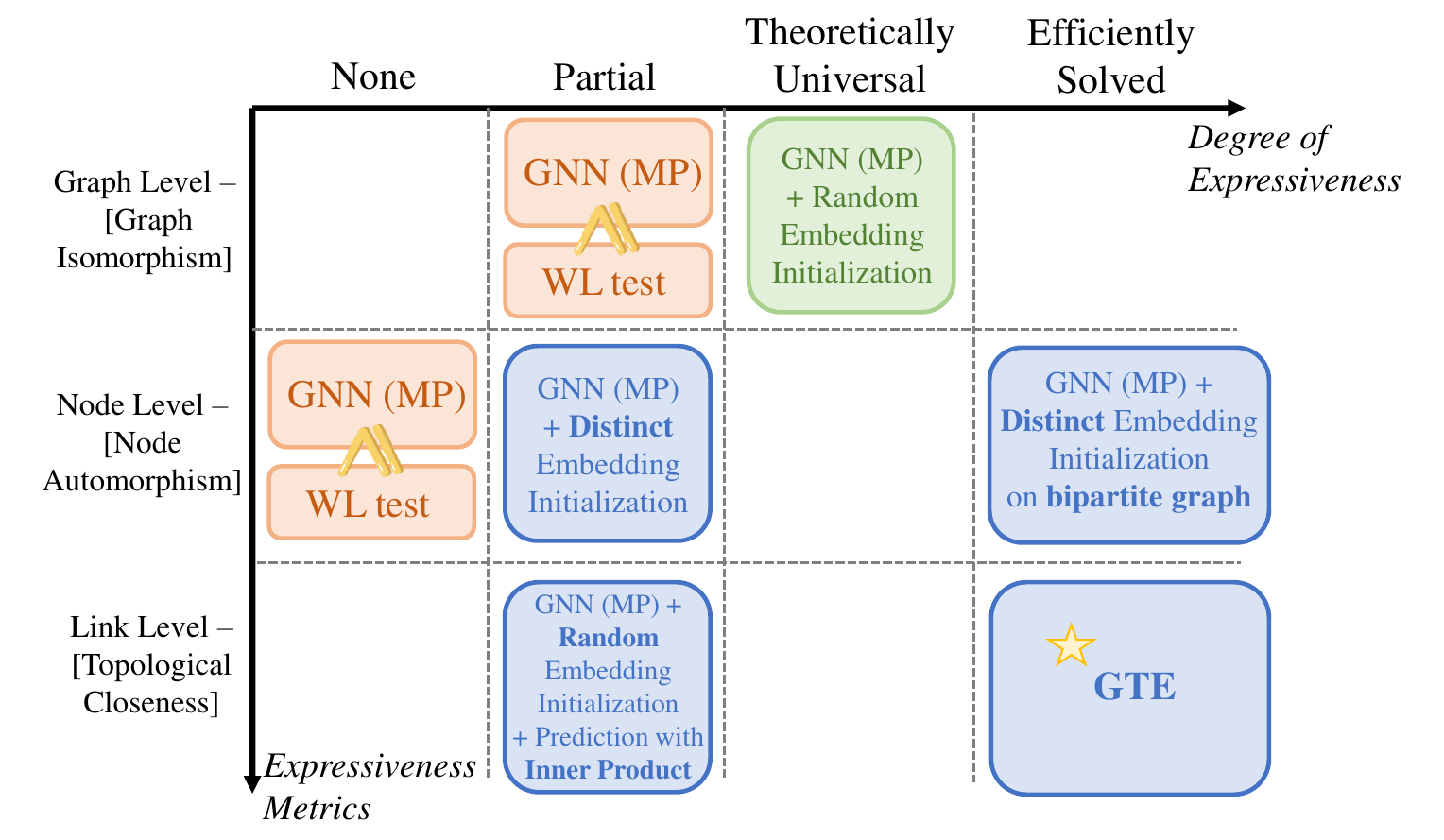}
            \subcaption{Current landscape of GNN expressiveness theorems (in light orange and green; "MP" denotes "Message Passing") and theorems proved in this paper (highlighted in blue).}\label{fig:overview12}
        \end{subfigure}
        \begin{subfigure}{\linewidth}
        \centering
            \includegraphics[width=0.7\textwidth]{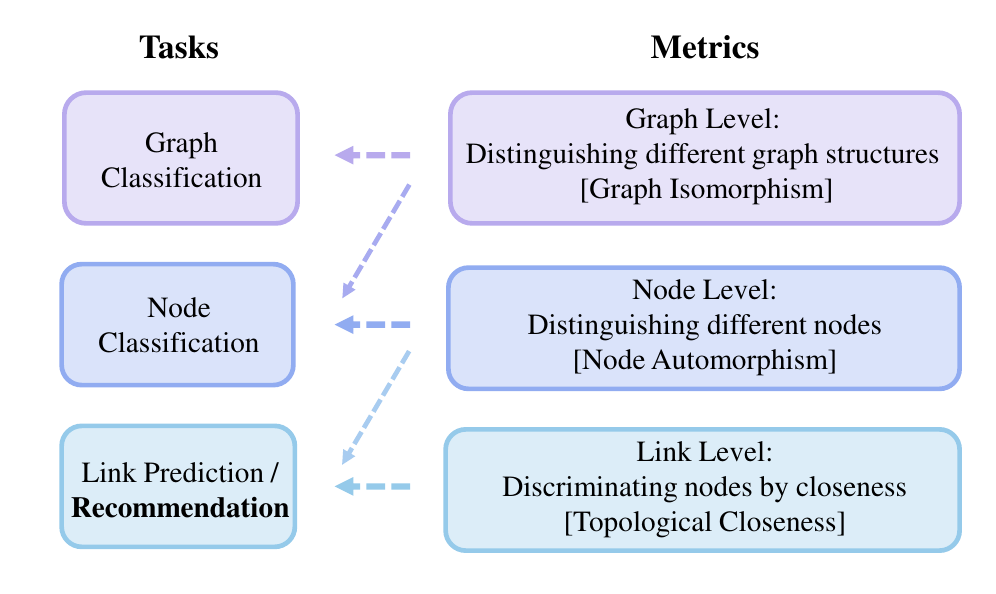}
            \subcaption{Relations between expressiveness metrics and tasks.}\label{fig:overview3}
        \end{subfigure}
        \vspace{-0.2in}
        \caption{Theoretical framework of GNN expressiveness.}\label{fig:overview}
        \vspace{-0.2in}
\end{figure} 

In recent years, there has been significant theoretical exploration into the expressive power of GNNs. Xu et al. \cite{xu2018powerful} proved that the message passing mechanism of GNNs is, at most, as expressive as the Weisfeiler-Lehman (WL) \cite{weisfeiler1968reduction} graph isomorphism test. Furthermore, it has been shown that initializing node features randomly can enhance the capability of GNNs and enable universal function approximation \cite{abboud2020surprising, sato2021random}. However, these theoretical findings have limited applicability to the recommendation task for two reasons. Firstly, the concept of "expressiveness" lacks a universal definition as it is strongly related to the specific task at hand. The aforementioned works utilize the graph isomorphism test as a metric of expressiveness. While this metric is relevant for graph-level tasks such as graph classification, it is less pertinent to recommendation, which primarily focuses on node similarity rather than graph-level properties. In their study on link prediction with subgraph sketching, Chamberlain et al. \cite{chamberlain2022graph} emphasize the importance of distinguishing automorphic nodes (symmetric nodes in the same orbit induced by the graph automorphism group) for link prediction, highlighting that the message passing mechanism of GNNs with equivalent power to the WL test lacks this ability. Although node automorphism is more relevant to link prediction and recommendation than graph isomorphism, it still does not fully align with the objective of recommendation, as it only requires distinguishing different nodes without considering their relative proximity. Secondly, existing works primarily focus on general graphs, while recommendation systems typically involve bipartite user-item interaction graphs, allowing for stronger conclusions regarding the expressiveness of GNNs (as we will demonstrate in Section \ref{sec:node}). Figure \ref{fig:overview12} provides an overview of the current progress in formulating the expressiveness of GNNs, depicted by the light orange and green boxes.

In this paper, we propose a comprehensive theoretical framework for analyzing the expressiveness of GNNs in recommendation, encompassing three levels of expressiveness metrics: i) Graph-level: The ability to distinguish isomorphic graphs, while less directly relevant to recommendation tasks, is included in our framework to ensure coherence and consistency with previous works~\cite{xu2018powerful,abboud2020surprising,sato2021random}. ii) Node-level: The ability to distinguish automorphic nodes, as mentioned in \cite{chamberlain2022graph}, is particularly relevant to recommendation systems as it assesses the model's capability to identify different users and items. At this level, we investigate the impact of equipping message passing GNNs with distinct (yet non-random) initial embeddings. Our research demonstrates that GNNs with distinct initial embeddings can successfully differentiate some of the automorphic nodes, although not all of them. Notably, we establish that when the graph is constrained to a bipartite structure, GNNs with distinct initial embeddings are capable of distinguishing all automorphic nodes. iii) Link-level: The ability to discriminate nodes of different topological closeness to a given node. We define topological closeness in Section \ref{sec:topo} and propose it as a new metric of expressiveness that directly aligns with the recommendation objective. Our theoretical analysis shows that the popular paradigm adopted in most GNN-based recommender systems (message passing GNN with random initial embeddings, using the inner product between embeddings as the prediction score) cannot fully discriminate nodes based on topological closeness. The relations between the three levels of metrics to different graph tasks are illustrated in Figure \ref{fig:overview3}.

It is worth noting that no single expressiveness metric can fully explain recommendation, as user preferences involve much more complicated factors than what can be encoded in a user-item interaction graph. Even if a metric directly aligns with the recommendation objective, achieving optimality on that metric does not guarantee flawless recommendation performance. Therefore, in Section \ref{sec:exp}, we analyze the effectiveness of topological closeness, the newly proposed metric, in explaining recommendation performance. Specifically, we propose a lightweight  \underline{G}raph \underline{T}opology \underline{E}ncoder (GTE) that adopts the message passing framework, but does not have learnable parameters. The learning-less characteristic of GTE makes it much more efficient than learning-based GNN recommenders. We prove that GTE is optimal on the new topological closeness metric and can achieve optimality on the node automorphism metric and the expressive power equivalent to the WL test on the graph isomorphism metric with suitable modification of the mapping function. Since GTE is optimal in discriminating nodes by topological closeness, we conduct various experiments with GTE and state-of-the-art GNN and GCL models to evaluate the effectiveness of the new metric in the recommendation task. The theories we prove in this paper and their relations to previous works are presented in Figure \ref{fig:overview12} (highlighted blue boxes).

In summary, our contributions are highlighted as follows:

\begin{itemize}[leftmargin=*]

\item We perform a comprehensive theoretical analysis on the expressiveness of GNNs in recommendation under a three-level framework designed specifically for the recommendation task.

\item We introduce a new link-level metric of GNN expressiveness, topological closeness, that directly aligns with the recommendation objective and is more suitable for evaluating recommendation expressiveness.

\item We propose a learning-less GNN algorithm GTE that is optimal on the link-level metric, whose learning-less feature enables it to be much more efficient than learning-based GNN recommenders.

\item We conduct extensive experiments on six real-world datasets of different sparsity levels with GTE and various baselines to explore the effectiveness of topological closeness in recommendation.

\end{itemize}

\section{Preliminaries and Related Work}

\subsection{GNNs for Recommendation}

The ability to extract multi-hop collaborative signals by aggregating neighborhood representation makes graph neural networks a prominent direction of research in recommender systems~\cite{wang2019neural,li2023graph}. Most GNN-based recommender models adopt the message passing type of GNNs, or more specifically, the graph convolutional networks (GCN) \cite{kipf2016semi}, as the backbone, such as NGCF \cite{wang2019neural} and PinSage \cite{ying2018graph}. GCCF \cite{chen2020revisiting} incorporates the residual structure into this paradigm. LightGCN \cite{he2020lightgcn} further adapts to the recommendation task by removing the non-linear embedding transformation. There are also attempts to enhance the GCN framework with supplementary tasks, such as masked node embedding reconstruction in \cite{zhang2019star}.

Recently, the self-supervised learning (SSL) paradigm has been incorporated into GNN-based recommenders to address the data sparsity issue \cite{wu2021self, yu2021self, xia2022self, wei2023multi}. The graph contrastive learning (GCL) is one of the most prominent lines of research that leverages the self-supervision signals produced by aligning embeddings learned from different views. Various GCL approaches have been proposed to produce augmented views, such as random feature dropout \cite{zhu2020deep}, hypergraph global learning \cite{xia2022hypergraph}, adaptive dropping based on node centrality \cite{zhu2021graph}, embeddings noisy perturbation \cite{yu2022graph}, and SVD-reconstruction \cite{cai2023lightgcl}. Despite their promising performance on real data, these models are often validated by experiments, with little theoretical formulation and guarantee on their capability.

\subsection{The Graph-Level Expressive Power of GNNs} \label{sec:graphlevel}

In recent years, there has been significant research investigating the expressiveness of GNNs using graph-level metrics, such as the graph isomorphism test. In particular, Xu et al. \cite{xu2018powerful} demonstrate that the expressive power of message passing GNNs is at most equivalent to the Weisfeiler-Lehman (WL) test, a popular graph isomorphism algorithm. Additionally, they establish that this equivalence can be achieved by ensuring that both the aggregation function and the readout function are injective. In our analysis, we adopt and adapt their conclusions, introducing relevant concepts and notations that will be utilized throughout our study.

\medskip
\textbf{Theorem 1} (from Xu et al. \cite{xu2018powerful}). \textit{Let $h_v^{(k)}$ be the feature of node $v$ in the $k$-th iteration, and $\mathcal{N}(v)$ be the set of neighboring nodes of v. With sufficient layers, a GNN can map any two graphs $G_1$ and $G_2$ that the WL-test decides as non-isomorphic to two different multisets of node features $\{h_v^{(k)}\}_1$ and $\{h_v^{(k)}\}_2$, if the aggregation function $h_v^{(k)} = g(\{h_u^{(k-1)}: u \in \mathcal{N}(v)\cup \{v\}\})$ is injective. Here, a multiset is defined as a set that allows multiple instances from its elements.}

\medskip


It is noteworthy that the aforementioned theorem solely focuses on the message passing mechanism of GNNs and assumes that nodes possess identical constant initial features, similar to the WL test. However, recent studies by Abboud et al~\cite{abboud2020surprising} and Sato et al~\cite{sato2021random} highlight that the capacity of GNNs can be significantly enhanced when nodes are equipped with randomly initialized features. Remarkably, random node initialization empowers GNNs with universality, enabling them to approximate any function on a graph and potentially solve the graph isomorphism problem~\cite{abboud2020surprising}. It is important to emphasize that this conclusion does not come with a complexity guarantee since the computational complexity of the graph isomorphism problem remains unknown. Random initialization represents a stronger assumption and presents greater challenges for analysis. In our paper, we specifically investigate the impact of a weaker assumption, distinct initialization, on the expressiveness of GNNs.


However, when evaluating the expressiveness of GNNs in recommendation systems, it is important to use metrics that align with the specific task of recommendation itself. Graph isomorphism tests and graph-level metrics, such as graph classification, may not directly capture the ability of GNNs to distinguish between different item nodes and accurately discriminate their closeness to a user node. In the context of link prediction, Chamberlain et al. \cite{chamberlain2022graph} suggest that the performance of message passing GNNs is hindered by their inability to differentiate automorphic nodes. Automorphic nodes refer to symmetric nodes within the same orbit induced by the graph automorphism group (e.g., nodes $a$ and $b$, $c$ and $d$, $e$ and $f$). The issue arises because automorphic nodes are assigned the same final features during the Weisfeiler-Lehman (WL) test, leading to indistinguishable effects in message passing GNNs. However, it is worth noting that this limitation assumes that all nodes have identical initial features. As a node-level metric, the ability to distinguish automorphic nodes becomes more relevant as it reflects the model's capability to differentiate between users and items. Therefore, it serves as a more appropriate metric for assessing GNN expressiveness in recommendation systems. Subsequent sections of our paper will demonstrate how GNNs, with distinct initial embeddings, can effectively address the issue of distinguishing automorphic nodes in user-item bipartite graphs.
\section{Expressing Node-Level Distinguishability of GNNs}
\label{sec:node}

\begin{figure}[t]
        \centering
            \includegraphics[width=0.7\linewidth]{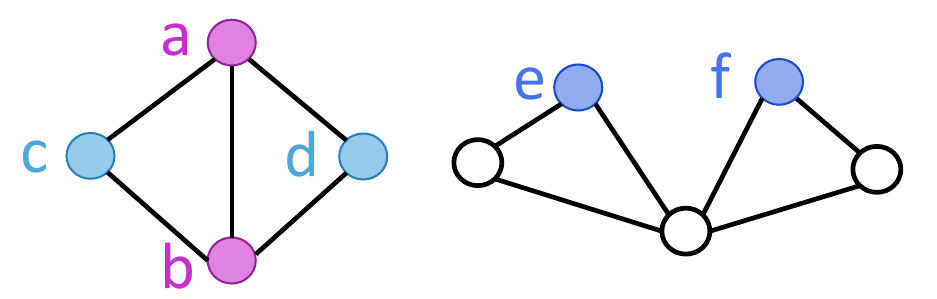}
            \caption{Illustrated examples of three types of automorphic nodes.($a$ and $b$ belong to Type II, $c$ and $d$ belong to Type I, $e$ and $f$ belong to Type III.)}
            \label{fig:auto}
\end{figure}

This section explores how distinct initial embeddings (DIE) enhance GNNs' ability to distinguish automorphic nodes. It's important to note that distinct initialization is a more relaxed assumption than random initialization in \cite{abboud2020surprising,sato2021random}, as it accepts any set of predefined unique initial embeddings, including the simplest one-hot encoding by node ID. We classify the automorphic nodes into three types:

\begin{itemize}[leftmargin=*]

\item \textbf{Type I}: a pair of automorphic nodes $u$ and $v$ such that $\mathcal{N}(u) = \mathcal{N}(v)$ (i.e., they share the same set of neighboring nodes). For example, $c$ and $d$ in Figure \ref{fig:auto}.

\item \textbf{Type II}: a pair of automorphic nodes $u$ and $v$ such that $u \in \mathcal{N}(v)$, $v \in \mathcal{N}(u)$, and $\mathcal{N}(u) - \{v\} = \mathcal{N}(v) - \{u\}$ (i.e., they are neighbors to each other, and share the same set of other neighboring nodes). For example, $a$ and $b$ in Figure \ref{fig:auto}.

\item \textbf{Type III}: a pair of automorphic nodes $u$ and $v$ such that $\mathcal{N}(u) - \{v\} \neq \mathcal{N}(v) - \{u\}$ (i.e., no matter if they are neighbors to each other, their neighborhoods differ by at least a pair of automorphic nodes $w$ and $w'$, such that $w \in {N}(u)$ but $w \notin {N}(v)$, $w' \in {N}(v)$ but $w' \notin {N}(u)$ \footnote{The reason why their neighborhoods must differ by at least a pair of nodes instead of one node is that in order for two nodes to be automorphic, their neighborhoods must have equal cardinality. The reason why the pair must be automorphic is that in order for two nodes to be automorphic, their rooted subtree structure must be the same. Or more intuitively, the neighborhoods of two symmetic nodes must also be symmetric.}). For example, $e$ and $f$ in Figure \ref{fig:auto}.

\end{itemize}

\subsection{DIE Does Not Fully Solve Node Automorphism on General Graphs}

In this section, we show that GNNs with distinct initial embeddings can only distinguish two of the three types of automorphic nodes.

\medskip
\textbf{Theorem 2}. \textit{Assuming that the aggregation function $g$ of the graph neural networks is injective, and every node receives a distinct initial embedding, for a pair of automorphic nodes $u$ and $v$, in every iteration, they will be assigned different embeddings if and only if one of the two conditions is satisfied:} \textit{(i) The GNN implements residual connections, and $u$ and $v$ belong to Type I or Type III.} \textit{(ii) The GNN does not implement residual connections, and $u$ and $v$ belong to Type II or Type III.} \textit{In other words, regardless of whether the GNN implements residual connections or not, it can only distinguish two out of the three types of automorphic nodes.}


\medskip
\textbf{Proof}. We first prove case (i), i.e, with residual connections:

We prove that for a pair of automorphic nodes $u$ and $v$ of Type I or III, they will be assigned different embeddings in any iteration $k$. This is obviously true when $k=0$, because the nodes are initialized with distinct embeddings. If it holds for iteration $k = i-1$, then in iteration $k=i$, the embeddings for $u$ and $v$ are:
\begin{equation}
    h_u^{(i)} = g(\{h_u^{(i-1)}\}\cup \{h_w^{(i-1)}:w\in \mathcal{N}(u)\})
\end{equation}

\begin{equation}
    h_v^{(i)} = g(\{h_v^{(i-1)}\}\cup \{h_{w'}^{(i-1)}:w'\in \mathcal{N}(v)\})
\end{equation}

If $u$ and $v$ are of Type I, then $\{h_u^{(i-1)}\}\cup \{h_w^{(i-1)}:w\in \mathcal{N}(u)\}$ and $\{h_v^{(i-1)}\}\cup \{h_{w'}^{(i-1)}:w'\in \mathcal{N}(v)\}$ are two distinct multisets, one exclusively containing $h_u^{(i-1)}$ and the other exclusively containing $h_v^{(i-1)}$, and $h_u^{(i-1)} \neq h_v^{(i-1)}$. If $u$ and $v$ are of Type III, the above-mentioned two multisets are also distinct, because there exists at least a pair of automorphic nodes $w$ and $w'$ (distinct from $u$, $v$), such that $w \in {N}(u)$ but $w \notin {N}(v)$, $w' \in {N}(v)$ but $w' \notin {N}(u)$. Thus, $h_w^{(i-1)}$ and $h_{w'}^{(i-1)}$ are exclusively in one of the two multisets, respectively. Moreover, $w$,$w'$ must be Type III automorphic nodes, because their neighborhoods differ by nodes other than each other ($u$ and $v$). By induction assumption, $h_w^{(i-1)} \neq h_{w'}^{(i-1)}$. In summary, the input multisets of $u$ and $v$ to $g$ must be different if they are Type I or III. Then, since $g$ is injective, $h_u^{(i)} \neq h_v^{(i)}$. By induction, this holds for all iterations. If $u$ and $v$ are of Type II, the inputs to the two multisets are the same, and the above result does not hold.

Next, we prove case (ii), i.e, without residual connection:


Similarly, we can demonstrate that for a pair of automorphic nodes $u$ and $v$ of Type II or III, they will be assigned different embeddings in any iteration $k$. This is evident when $k=0$ because the nodes are initially initialized with distinct embeddings. Assuming it holds true for iteration $k=i-1$, we can show that when $k=i$, the embeddings for nodes $u$ and $v$ are as follows:
\begin{equation}
    h_u^{(i)} = g(\{h_w^{(i-1)}:w\in \mathcal{N}(u)\})
\end{equation}
\begin{equation}
    h_v^{(i)} = g(\{h_{w'}^{(i-1)}:w'\in \mathcal{N}(v)\})
\end{equation}


If nodes $u$ and $v$ are of Type II, we can observe that the multisets ${h_w^{(i-1)} : w \in \mathcal{N}(u)}$ and ${h_{w'}^{(i-1)} : w' \in \mathcal{N}(v)}$ are two distinct multisets. Specifically, one exclusively contains the embedding $h_v^{(i-1)}$, while the other exclusively contains the embedding $h_u^{(i-1)}$. It is also established that $h_u^{(i-1)} \neq h_v^{(i-1)}$. if nodes $u$ and $v$ are of Type III, the arguments presented in the case of Type II still apply. Thus, the input multisets of nodes $u$ and $v$ to the aggregation function $g$ must be different if they are of Type I or II. Then, since the aggregation function $g$ is injective, we can conclude that $h_u^{(i)} \neq h_v^{(i)}$. By induction, we can assert that this holds for all iterations.

However, if $u$ and $v$ are of Type I, the inputs to the two multisets are the same, and the above result does not hold. \hfill $\square$

\medskip

This theorem suggests that on a general graph, using distinct embedding initialization can only provide partial differentiation between automorphic nodes for GNNs. However, it is important to note that user-item interaction graphs in recommendation systems are bipartite graphs, where a user node cannot be connected to another user node, and an item node cannot be connected to another item node. This bipartite structure allows us to leverage this inherent feature and demonstrate that with distinct initial embeddings, GNNs can fully address the node automorphism problem on user-item bipartite graphs.

\subsection{DIE Solves Node Automorphism on Bipartite Graphs} \label{sec:diebipartite}

We start by proving that Type II automorphic nodes cannot exist in a connected bipartite graph with more than two nodes.

\medskip
\textbf{Lemma 3}. \textit{In a connected bipartite graph with more than 2 nodes, if a pair of nodes $u$ and $v$ are automorphic, they cannot be Type II automorphic nodes.}

\medskip
\textbf{Proof}. If $u$ and $v$ are both user nodes or both item nodes, there cannot be an edge between them, so the condition for Type II, $u \in \mathcal{N}(v)$ and $v \in \mathcal{N}(u)$, cannot hold. If one of $u$, $v$ is a user node and another one is an item node, and they are neighbors to each other, then since all neighbors of a user node must be item nodes, and all neighbors of an item node must be user nodes, $\mathcal{N}(u) - \{v\}$ cannot be equal to $\mathcal{N}(v) - \{u\}$, unless $\mathcal{N}(u) - \{v\} = \emptyset$ and $\mathcal{N}(v) - \{u\} = \emptyset$. If $\mathcal{N}(u) - \{v\} = \emptyset$ and $\mathcal{N}(v) - \{u\} = \emptyset$, then $u$ and $v$ must form a connected component that is isolated from the rest of the graph. However, since the graph is a connected bipartite graph with more than 2 nodes, the above scenario cannot occur. \hfill $\square$

\medskip
With Lemma 3 and Theorem 2, we prove the following theorem.

\medskip
\textbf{Theorem 4}. \textit{Assume that the graph is a connected bipartite graph with more than 2 nodes. Assume that the aggregation function $g$ of the GNN is injective, and residual connections are implemented. Assume that every node receives a distinct initial embedding. For a pair of automorphic nodes $u$ and $v$, in every iteration, they will be assigned different embeddings.}

\medskip
\textbf{Proof}. By Lemma 3, $u$ and $v$ must be either Type I or III automorphic nodes. By Theorem 2, as long as the GNN implements residual connections, $u$ and $v$ will be assigned different embeddings in every iteration due to the injectivity of the aggregation function. \hfill $\square$

\medskip
This conclusion indicates that a GNN is capable of distinguishing automorphic nodes on a connected user-item bipartite graph with the following three design choices: i) distinct initial embeddings; ii) residual connections; iii) injective aggregation function. These design choices provide a useful guideline for developing powerful GNN-based recommender systems that aim to achieve optimal expressiveness on the node-level metric.
\section{Encoding Node Topological Closeness with GNNs}
\label{sec:topo}


While the capability of node automorphism provides a better metric for recommendation expressiveness, it is not directly aligned with the intended task. This is because it only focuses on distinguishing between different nodes, without considering their structural closeness to one another. In recommendation systems, models should not only differentiate between two items but also determine which one is closer to the target user's profile. Therefore, it is important to focus on a link-level metric that captures the closeness between nodes in the graph's structural space.


The traditional metric for evaluating the closeness between two nodes is the geodesic distance, which is defined as the length of the shortest path between the two nodes~\cite{graphtheory,bouttier2003geodesic}. However, this metric may not accurately reflect the true similarity between nodes as encoded in the graph structure. This is especially true in tasks where clusters and community structures are important, such as recommendation systems. For instance, in Figure \ref{fig:tc2}, nodes $v_1$ and $v_2$ have the same geodesic distance to node $u$. However, it is evident that $v_2$ should be considered "closer" to $u$ than $v_1$ because they reside together in a denser cluster.

In light of this limitation, we propose a new link-level metric \textit{topological closeness} that evaluates the closeness between two nodes by considering not only the length of the shortest path but also the number of possible paths between them.

\subsection{Topological Closeness}

Here, we define the k-hop topological closeness between two nodes $u$ and $v$, denoted as $k\mbox{-}TC(u,v)$, as follows.

\medskip
\textbf{Definition 5} (k-Hop Topological Closeness). \textit{Given two nodes $u$ and $v$ in an undirected graph, the k-hop topological closeness between $u$ and $v$ in this graph is defined as:}
\begin{equation*}
    k\mbox{-}TC(u,v) = \left| \mathcal{P}_{u,v}^{k} \right|
\end{equation*}

\noindent
\textit{where $\mathcal{P}_{u,v}^{k}$ is the set of all possible paths of length $k$ between $u$ and $v$. Note that the paths here allow repeated vertices, repeated edges, and self-loops. We further define the 0-hop topological distance between a node $u$ and itself as $0\mbox{-}TC(u,u)=1$.}

 \begin{figure}[t]
         \begin{subfigure}{\linewidth}
         \centering
             \includegraphics[width=0.7\textwidth]{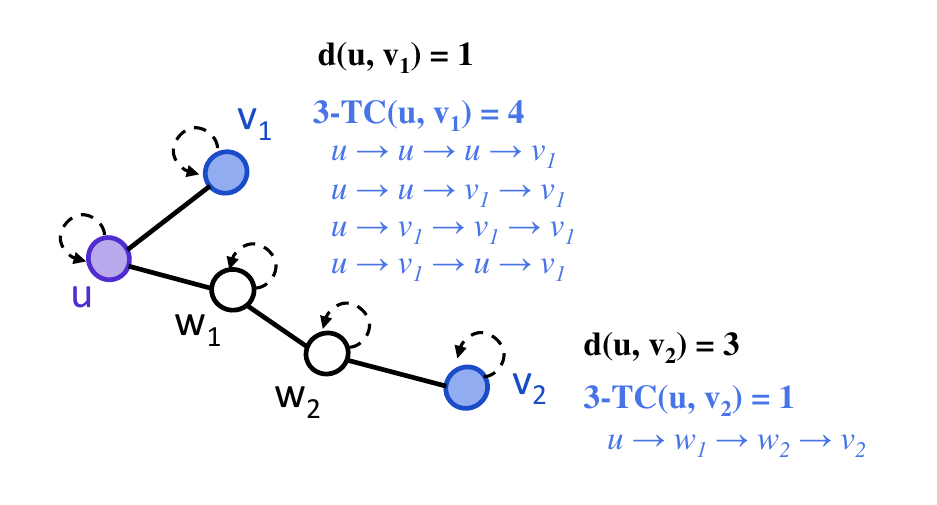}
             \subcaption{$v_1$ and $v_2$ have one simple path to $u$ but with different distance.}\label{fig:tc1}
         \end{subfigure}
         \begin{subfigure}{\linewidth}
         \centering
             \includegraphics[width=0.7\textwidth]{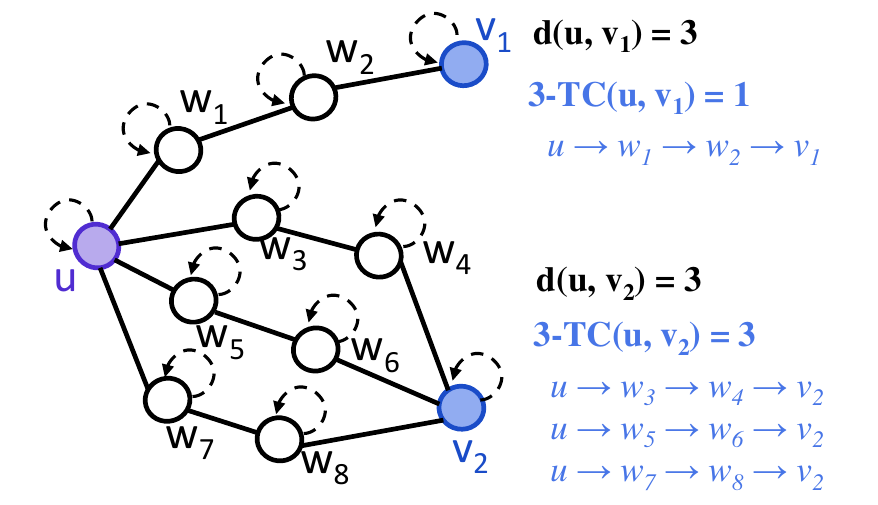}
             \subcaption{$v_1$ and $v_2$ have the same shortest distance to $u$ but with different clustering patterns.}\label{fig:tc2}
         \end{subfigure}
         \caption{Illustrations of Topological Closeness with k=3.}\label{fig:tc}
         \vspace{-0.15in}
 \end{figure}

\medskip
The capability of topological closeness (TC) to intuitively capture both the distance and clustering information between two nodes is illustrated in Figure \ref{fig:tc}. In Figure \ref{fig:tc1}, both $v_1$ and $v_2$ have only one simple path\footnote{A simple path is defined as a path in which all the internal vertices are distinct.} to $u$, representing the minimum clustering structure they can have when connected to $u$. Additionally, $v_1$ has a shorter distance to $u$ compared to $v_2$. Consequently, $v_1$ exhibits a higher TC value to $u$ than $v_2$. This occurs because TC takes into account self-loops and repeated vertices, and a shorter simple path allows more possible paths of length $k$ with self-loops and repeated vertices. In Figure \ref{fig:tc2}, both $v_1$ and $v_2$ are equidistant from $u$. However, $v_2$ resides within a denser cluster alongside $u$. Thus, $v_2$ demonstrates a higher TC value to $u$ compared to $v_1$, due to the larger number of paths connecting $v_2$ and $u$ than those connecting $v_1$ and $u$.



It is important to note that there are multiple methods available for combining clustering density information and the shortest distance between two nodes. For instance, one approach involves calculating a weighted sum of all simple paths connecting the two nodes, where the weight is determined by the inverse of the path length. In our case, we define topological closeness in this manner because this definition aligns naturally with the structure of GNNs and can be efficiently computed using a message-passing GNN. The subsequent two subsections will provide further clarity on this concept and its relevance within our framework.

\subsection{Popular GNN Recommender Paradigm Does Not Fully Capture Topological Closeness} \label{sec:theoGNNtopo}


In recent years, most GNN-based recommender systems \cite{he2020lightgcn,wu2021self,xia2022hypergraph,yu2022graph,cai2023lightgcl} adopt the following paradigm. Initially, each user $u$ and item $v$ are assigned random initial embedding vectors $\mathbf{e}_u^{(0)} \in \mathbb{R}^d$ and $\mathbf{e}_v^{(0)} \in \mathbb{R}^d$ respectively, where $d$ represents the dimensionality of the embeddings. In the $k$-th GNN layer, the embeddings are updated using the following rule, which incorporates residual connections:

\begin{equation}
    \mathbf{e}_u^{(k)} = \sigma(\smashoperator[r]{\sum_{v \in \mathcal{N}(u)}} \phi(\mathbf{e}_v^{(k-1)})) + \mathbf{e}_u^{(k-1)}
\end{equation}

\begin{equation}
    \mathbf{e}_v^{(k)} = \sigma(\smashoperator[r]{\sum_{u \in \mathcal{N}(v)}} \phi(\mathbf{e}_u^{(k-1)})) + \mathbf{e}_v^{(k-1)}
\end{equation}

\noindent
where $\sigma$ denoting an activation function and $\phi$ generally representing the element-wise multiplication with the coefficients in the normalized adjacency matrix, the final embeddings of the nodes can be computed as follows:


\begin{equation}
    \mathbf{e}_u = \theta(\{\mathbf{e}_u^{(k)}\}_{k=0}^L), \quad \mathbf{e}_v = \theta(\{\mathbf{e}_v^{(k)}\}_{k=0}^L)
\end{equation}

\noindent
where $L$ represents the number of GNN layers, and $\theta$ denotes an aggregation function applied to the embeddings from all layers, typically implemented as a sum or mean operation. Finally, the predicted score of a user $u$'s preference for an item $v$ is computed by taking the inner product of their final embeddings:

\begin{equation}
    \hat{y}_{u,v} = \mathbf{e}_u \cdot \mathbf{e}_v
\end{equation}


This section provides a proof demonstrating that the message passing GNN and inner product prediction mechanism employed in this paradigm are unable to completely discriminate between two item nodes based solely on their topological distance from a user node. To further substantiate this claim, we conduct experiments on real-world datasets, which are presented in Section \ref{sec:topoGNNexp}.

For the sake of simplicity in calculations and clearer presentation, we consider a simplified version of the aforementioned paradigm. Specifically, we assume that $\sigma$ and $\phi$ are identity functions. Hence, the embedding updating rule is simplified as:
\begin{equation}
    \mathbf{e}_u^{(k)} = \smashoperator[r]{\sum_{v \in \mathcal{N}(u)}} \mathbf{e}_v^{(k-1)} + \mathbf{e}_u^{(k-1)}, \ \ \mathbf{e}_v^{(k)} = \smashoperator[r]{\sum_{u \in \mathcal{N}(v)}} \mathbf{e}_u^{(k-1)} + \mathbf{e}_v^{(k-1)}
\end{equation}


Furthermore, we make the additional assumption that the embedding from the $L$-th layer, which is the final layer, is utilized as the final embedding for the computation of the inner product.
\begin{equation}
    \hat{y}_{u,v} = \mathbf{e}_u^{(L)} \cdot \mathbf{e}_v^{(L)}
\end{equation}

With this simplified paradigm, we show that when an item $v_1$ has higher topological closeness to a user $u$ than another item $v_2$, the prediction score $\hat{y}_{u,v_1}$ is not necessarily higher than $\hat{y}_{u,v_2}$.


\medskip
\textbf{Lemma 6}. \textit{With the simplified paradigm described above, in layer $k$, the embedding of a user $u$ can be written as follows:}

\begin{equation*}
    \mathbf{e}_u^{(k)} = \smashoperator[r]{\sum_{w \in \mathcal{N}^{(k)}(u) \cup \{u\}}} k\mbox{-}TC(u, w) \cdot \mathbf{e}_w^{(0)}
\end{equation*}

\noindent
\textit{where $\mathcal{N}^{(k)}(u)$ is the set of k-hop neighbors of $u$ (i.e., all nodes whose distance to $u$ is smaller than or equal to k). Specially, $\mathcal{N}(u) = \mathcal{N}^{(1)}(u)$ is the set of (1-hop) neighbors of $u$.}

\medskip
\textbf{Proof}. This is obviously true when $k = 0$, because 
\begin{equation*}
    \smashoperator[r]{\sum_{w \in \mathcal{N}^{(0)}(u) \cup \{u\}}} 0\mbox{-}TC(u, w) \cdot \mathbf{e}_w^{(0)} = 0\mbox{-}TC(u, u) \cdot \mathbf{e}_u^{(0)} = \mathbf{e}_u^{(0)}
\end{equation*}

If it holds for $k = l - 1$, then in layer $k = l$, the embedding of $u$ is:
\begin{align}
    \mathbf{e}_u^{(l)} &= \sum_{v \in \mathcal{N}(u)}\mathbf{e}_v^{(l-1)} + \mathbf{e}_u^{(l-1)} \\
    &= \sum_{v \in \mathcal{N}(u) \cup \{u\}} \left( \smashoperator[r]{\sum_{w \in \mathcal{N}^{(l-1)}(v)\cup\{v\}}} (l-1)\mbox{-}TC(v, w) \cdot \mathbf{e}_w^{(0)} \right)\\
    &= \sum_{v \in \mathcal{N}(u) \cup \{u\}} \left( \smashoperator[r]{\sum_{w \in \mathcal{N}^{(l-1)}(v)\cup\{v\}}} \left| \mathcal{P}_{v, w}^{(l-1)} \right| \cdot \mathbf{e}_w^{(0)} \right)\\
    &= \sum_{w \in \mathcal{N}^{(l)}(u)\cup\{u\}} \left(\smashoperator[r]{\sum_{v \in \mathcal{N}(u) \cup \{u\}}} \left| \mathcal{P}_{v, w}^{(l-1)} \right| \right) \cdot \mathbf{e}_w^{(0)} \label{eq:emb}
\end{align}

Note that for every path from $w$ to $v$ with length $l-1$, it must correspond to a path from $w$ to $u$ with length $l$, with the second last node being $v$, because $v$ is a neighbor of $u$. Then, $\left| \mathcal{P}_{v, w}^{(l-1)} \right| = \left| \mathcal{P}_{u, w,\sim v}^{(l)} \right|$, where $\mathcal{P}_{u, w,\sim v}^{(l)}$ denotes the set of paths from $w$ to $u$ with length $l$, with the second last node being $v$. Further note that $\bigcup_{v \in \mathcal{N}(u) \cup \{u\}}\mathcal{P}_{u, w,\sim v}^{(l)} = \mathcal{P}_{u, w}^{(l)}$, because any path from $w$ to $u$ must have the second last node being one of $u$'s neighbors or $u$ itself (self-loop as the last edge). Also, for two distinct $v, v' \in \mathcal{N}(u) \cup \{u\}$, the sets $\mathcal{P}_{u, w,\sim v}^{(l)}$ and $\mathcal{P}_{u, w,\sim v'}^{(l)}$ are disjoint. So it follows that $\sum_{v \in \mathcal{N}(u) \cup \{u\}} \left| \mathcal{P}_{v, w, \sim v}^{(l)} \right| = \left| \mathcal{P}_{v, w}^{(l)} \right|$. With these conclusions, Expression (\ref{eq:emb}) can be further transformed into:
\begin{align}
    \mathbf{e}_u^{(l)} &= \sum_{w \in \mathcal{N}^{(l)}(u)\cup\{u\}} \left(\smashoperator[r]{\sum_{v \in \mathcal{N}(u) \cup \{u\}}} \left| \mathcal{P}_{u, w, \sim v}^{(l)} \right| \right) \cdot \mathbf{e}_w^{(0)} \\
    &= \smashoperator[r]{\sum_{w \in \mathcal{N}^{(l)}(u)\cup\{u\}}} \left| \mathcal{P}_{u, w}^{(l)} \right| \cdot \mathbf{e}_w^{(0)} \\
    &= \smashoperator[r]{\sum_{w \in \mathcal{N}^{(l)}(u) \cup \{u\}}} l\mbox{-}TC(u, w) \cdot \mathbf{e}_w^{(0)}
\end{align}
By induction, the expression holds for every iteration $k$.  \hfill $\square$

\medskip
With this Lemma, in the simplified paradigm described above, for a user node $u$ and two item nodes $v_1$ and $v_2$, if $L\mbox{-}TC(u,v_1) > L\mbox{-}TC(u,v_2)$, we explore the relationship between their respective predicted scores $\hat{y}_{u,v_1}$ and $\hat{y}_{u,v_2}$ after $L$ GNN layers:

\small
\begin{align*}
    & \hat{y}_{u,v_1} - \hat{y}_{u,v_2}
    = \mathbf{e}_u^{(L)} \cdot \mathbf{e}_{v_1}^{(L)} - \mathbf{e}_u^{(L)} \cdot \mathbf{e}_{v_2}^{(L)} \\
    &= \left( \smashoperator[r]{\sum_{w \in \mathcal{N}^{(L)}(u) \cup \{u\}}} L\mbox{-}TC(u, w) \cdot \mathbf{e}_w^{(0)} \right) \cdot \left( \smashoperator[r]{\sum_{w' \in \mathcal{N}^{(L)}(v_1) \cup \{v_1\}}} L\mbox{-}TC(v_1, w') \cdot \mathbf{e}_{w'}^{(0)} \right) \\
    &\ \ - \left( \smashoperator[r]{\sum_{w \in \mathcal{N}^{(L)}(u) \cup \{u\}}} L\mbox{-}TC(u, w) \cdot \mathbf{e}_w^{(0)} \right) \cdot \left( \smashoperator[r]{\sum_{w'' \in \mathcal{N}^{(L)}(v_2) \cup \{v_2\}}} L\mbox{-}TC(v_2, w'') \cdot \mathbf{e}_{w''}^{(0)} \right) \\
    &= \left( k_1 + L\mbox{-}TC(u, v_1) \cdot \mathbf{e}_{v_1}^{(0)} \right) \cdot \left( k_3 + L\mbox{-}TC(u, v_1) \cdot \mathbf{e}_{u}^{(0)} \right) \\
    &\ \ - \left( k_2 + L\mbox{-}TC(u, v_2) \cdot \mathbf{e}_{v_2}^{(0)} \right) \cdot \left( k_4 + L\mbox{-}TC(u, v_2) \cdot \mathbf{e}_{u}^{(0)} \right) \\
    &= (k_1k_3 - k_2k_4) \\
    &\ \ + \left( (k_1 \mathbf{e}_{u}^{(0)} + k_3 \mathbf{e}_{v_1}^{(0)}) \underline{L\mbox{-}TC(u, v_1)} - (k_2 \mathbf{e}_{u}^{(0)} + k_4 \mathbf{e}_{v_2}^{(0)}) \underline{L\mbox{-}TC(u, v_2)} \right) \\
    &\ \ + \left( \mathbf{e}_{u}^{(0)}\mathbf{e}_{v_1}^{(0)}\left( \underline{L\mbox{-}TC(u, v_1)} \right)^2 - \mathbf{e}_{u}^{(0)}\mathbf{e}_{v_2}^{(0)}\left( \underline{L\mbox{-}TC(u, v_2)} \right)^2 \right)
\end{align*}
\normalsize

\noindent
where
\noindent
\small
\begin{equation*}
    k_1 = \smashoperator[r]{\sum_{\substack{w \in \mathcal{N}^{(L)}(u) \cup \\ \{u\} - \{v_1\}}}} L\mbox{-}TC(u, w) \cdot \mathbf{e}_w^{(0)}, \quad k_2 = \smashoperator[r]{\sum_{\substack{w \in \mathcal{N}^{(L)}(u) \cup \\ \{u\} - \{v_2\}}}} L\mbox{-}TC(u, w) \cdot \mathbf{e}_w^{(0)},
\end{equation*}

\begin{equation*}
    k_3 = \smashoperator[r]{\sum_{\substack{w' \in \mathcal{N}^{(L)}(v_1) \cup \\ \{v_1\} - \{u\}}}} L\mbox{-}TC(v_1, w') \cdot \mathbf{e}_{w'}^{(0)}, \quad k_4 = \smashoperator[r]{\sum_{\substack{w'' \in \mathcal{N}^{(L)}(v_2) \cup \\ \{v_2\} - \{u\}}}} L\mbox{-}TC(v_2, w'') \cdot \mathbf{e}_{w''}^{(0)}
\end{equation*}
\normalsize

We can see that the expression of  $(\hat{y}_{u,v_1} - \hat{y}_{u,v_2})$ contains the difference between $L\mbox{-}TC(u, v_1)$ and $L\mbox{-}TC(u, v_2)$, but each with a coefficient that depends on other factors, such as the inner product between the randomly initialized embeddings and the graph structure involving other nodes. Even if $L\mbox{-}TC(u, v_1) > L\mbox{-}TC(u, v_2)$, it is uncertain whether $\hat{y}_{u,v_1} > \hat{y}_{u,v_2}$ or not. This indicates that the widely adopted paradigm of message passing GNN and inner product prediction for recommender systems cannot fully capture the topological closeness between the nodes. It is worth noting that we only study the model structure of message passing and prediction in the above analysis, without considering the learning process. It is possible that the models could fit the task better with learnable embeddings that are widely adopted in GNN recommenders, but theoretical analysis involving learning would require complicated case-by-case analysis that we cannot include here.

\subsection{Graph Topology Encoder (GTE)}

\small
\begin{algorithm}[t]
\caption{GTE}\label{alg:gte}
\KwInput{adjacency matrix of the user-item interaction graph $\mathcal{A} \in \mathbb{R}^{U \times I}$, where $U$ and $I$ are the number of users and items, respectively.}
\KwOutput{$\left\{\mathbf{h}_{u_i}^{(L)}\right\}_{i=0}^{U-1}$ s.t. prediction $\hat{y}_{i,j} = \mathbf{h}_{u_i}^{(L)}\lbrack j \rbrack$.}
Assign initial features $\left\{\mathbf{h}_{u_i}^{(0)} \in \mathbb{R}^I \right\}_{i=0}^{U-1}$ and  $\left\{\mathbf{h}_{v_j}^{(0)} \in \mathbb{R}^I \right\}_{j=0}^{I-1}$, let $\mathbf{H}_u^{(0)} \in \mathbb{R}^{U \times I}$ and $\mathbf{H}_v^{(0)} \in \mathbb{R}^{I \times I}$ be the collections of user and item initial features in the matrix form\;
\For{$k=1$ to $L$}{
$\mathbf{H}_u^{(k)} \leftarrow \mathcal{A} \cdot \mathbf{H}_v^{(k-1)} + \mathbf{H}_u^{(k-1)}$\;
$\mathbf{H}_v^{(k)} \leftarrow \mathcal{A}^\top \cdot \mathbf{H}_u^{(k-1)} + \mathbf{H}_v^{(k-1)}$\;
}
\end{algorithm}
\normalsize


Indeed, it is evident that no single metric can fully capture a model's capability in recommendation, even if the metric aligns with the recommendation objective. User behavior is considerably more intricate than what is encompassed within the user-item graph's interaction history. As a result, it is crucial to investigate the level of interpretability of the proposed metric in the context of recommendation. To address this concern, we introduce a learning-less GNN algorithm termed GTE. GTE is proven to be optimal in terms of the link-level metric of topological closeness.

At the beginning, each item $v_i$ is assigned a one-hot initial feature $\mathbf{h}_{v_i}^{(0)} \in \mathbb{R}^I$, where $I$ is the total number of items. For $j \in \lbrack 0, I)$, the $j$-th entry of the initial feature is \small $
    \mathbf{h}_{v_i}^{(0)}\lbrack j \rbrack = \begin{cases}
    1 & i=j\\
    0 & i\neq j\\
    \end{cases}
$ \normalsize.


Each user $u_i$ is assigned an initial feature vector $\mathbf{h}_{u_i}^{(0)} \in \mathbb{R}^I$, where all entries are initialized to 0. It is important to note that, unlike traditional GNN-based recommenders where the embeddings are learnable, the features in GTE are fixed. In fact, the entire algorithm does not involve any learning process. In the $k$-th GNN layer, the features are propagated on the graph using a simple rule:

\small
\begin{equation*}
    \mathbf{h}_u^{(k)} = \phi \left( \smashoperator[r]{\sum_{v \in \mathcal{N}(u)}} \mathbf{h}_v^{(k-1)} + \mathbf{h}_u^{(k-1)} \right),\ \  \mathbf{h}_v^{(k)} = \phi \left( \smashoperator[r]{\sum_{u \in \mathcal{N}(v)}} \mathbf{h}_u^{(k-1)} + \mathbf{h}_v^{(k-1)} \right)
\end{equation*}
\normalsize
\noindent
where $\phi$ is a mapping function, and in our case we simply use the identity function as $\phi$. The message propagation is performed $L$ times, where $L$ is the number of GNN layers. Finally, for user $u_i$, the predicted preference score for item $v_j$ is $\hat{y}_{i,j} = \mathbf{h}_{u_i}^{(L)}\lbrack j \rbrack$. Note that unlike most GNN-based recommenders that use inner product for prediction, GTE directly uses a specific dimension in the feature space to represent the score for an item.

The procedure of GTE is formally presented in Algorithm \ref{alg:gte}. GTE exploits the graph collaborative filtering signals via a fixed GNN structure without any learnable components, which makes the execution of the algorithm fast and reliable. We prove that GTE is optimal on the proposed topological closeness metric by showing that for a user node $u_i$ and two item nodes $v_j$ and $v_k$, if $L\mbox{-}TC(u_i,v_j) > L\mbox{-}TC(u_i,v_k)$, it is certain that $\hat{y}_{i,j} > \hat{y}_{i,k}$.

\medskip
\textbf{Lemma 7}. \textit{In the $k$-th layer of GTE, the $j$-th dimension of the feature of a node $w$ (it could be user or item node) can be written as follows, where $v_j$ is the $j$-th item node:}

\noindent
\begin{equation*}
    \mathbf{h}_{w}^{(k)}\lbrack j \rbrack = k\mbox{-}TC(w, v_j)
\end{equation*}

\medskip
\textbf{Proof}. This is true when $k=0$, because if $w \neq v_j$, $\mathbf{h}_{w}^{(0)}\lbrack j \rbrack = 0$, and $0\mbox{-}TC(w, v_j) = 0$; if $w = v_j$, $\mathbf{h}_{w}^{(0)}\lbrack j \rbrack = \mathbf{h}_{v_j}^{(0)}\lbrack j \rbrack = 1$, and $0\mbox{-}TC(w, v_j) = 0\mbox{-}TC(v_j, v_j) = 1$.

If it holds for $k = l - 1$, then in layer $k = l$, the $j$-th entry of the feature of $w$ is:
\small
\begin{align}
    \mathbf{h}_w^{(l)} \lbrack j \rbrack &= \sum_{x \in \mathcal{N}(w)}\mathbf{h}_x^{(l-1)} \lbrack j \rbrack + \mathbf{h}_w^{(l-1)} \lbrack j \rbrack \\ &= \smashoperator[r]{\sum_{x \in \mathcal{N}(w)\cup\{w\}}} (l-1)\mbox{-}TC(x,v_j) \\
    &= \smashoperator[r]{\sum_{x \in \mathcal{N}(w)\cup\{w\}}} \left| \mathcal{P}_{x,v_j}^{(l-1)} \right| \\ &= \smashoperator[r]{\sum_{x \in \mathcal{N}(w)\cup\{w\}}} \left| \mathcal{P}_{w,v_j,\sim x}^{(l)} \right| \\
    &= \left| \mathcal{P}_{w,v_j}^{(l)} \right| = l\mbox{-}TC(w,v_j)
\end{align}
\normalsize

By induction, the expression holds for every iteration k.  \hfill $\square$

\medskip
\textbf{Theorem 8}. \textit{In GTE, after $L$ layers, for a user node $u_i$ and two item nodes $v_j$ and $v_k$, if $L\mbox{-}TC(u_i,v_j) > L\mbox{-}TC(u_i,v_k)$, then $\hat{y}_{i,j} > \hat{y}_{i,k}$.}

\medskip
\textbf{Proof}. We have shown in Lemma 7 that $\hat{y}_{i,j} = \mathbf{h}_{u_i}^{(L)}\lbrack j \rbrack = L\mbox{-}TC(u_i, v_j)$, and $\hat{y}_{i,k} = \mathbf{h}_{u_i}^{(L)}\lbrack k \rbrack = L\mbox{-}TC(u_i, v_k)$. Thus, if it holds that $L\mbox{-}TC(u_i,v_j) > L\mbox{-}TC(u_i,v_k)$, then $\hat{y}_{i,j} > \hat{y}_{i,k}$.  \hfill $\square$

\medskip
The above analysis proves that GTE is optimal on the link-level metric topological closeness, and thus its performance on recommendation datasets can be used to evaluate the effectiveness of the proposed metric.

\begin{table*}[t]
    \centering
    \caption{Performance of GTE and baselines on datasets of different data density, in terms of Recall (R) and NDCG (N).}
    \label{tab:result}
    \small
    \begin{tabular}{cccccccccccccc}
        \toprule
        Dataset & Density & Metric & BiasMF & NGCF & GCCF & LightGCN & SimGRACE & SAIL & HCCF & LightGCL & \underline{GTE} & \textit{p-val} & \textit{impr.}\\
        \midrule
        \multicolumn{14}{l}{\textit{Denser datasets}} \\
        \midrule
        \multirow{4}{*}{Yelp} & \multirow{4}{*}{$2.1e^{-3}$} & R@20 & 0.0190 & 0.0294 & 0.0462 & 0.0482 & 0.0603 & 0.0471 & 0.0626 & \textbf{0.0793} & 0.0647 & $3e^{-13}$ & -18\% \\
        &&N@20 & 0.0161 & 0.0243 & 0.0398 & 0.0409 & 0.0435 & 0.0405 & 0.0527 & \textbf{0.0668} & 0.0556 & $3e^{-14}$ & -16\% \\
        &&R@40 & 0.0371 & 0.0522 & 0.0760 & 0.0803 & 0.0989 & 0.0733 & 0.1040 & \textbf{0.1292} & 0.0556 & $2e^{-14}$ & -18\% \\
        &&N@40 & 0.0227 & 0.0330 & 0.0508 & 0.0527 & 0.0656 & 0.0516 & 0.681 & \textbf{0.0852} & 0.0704 & $4e^{-15}$ & -17\% \\
         \midrule
         &  & R@20 & 0.0926 & 0.0999 & 0.0986 & 0.1006 & 0.0827 & 0.1211 & 0.1083 & 0.1216 & \textbf{0.1269} & $3e^{-7}$ & 4\% \\
 &  & N@20 & 0.0687 & 0.0739 & 0.0730 & 0.0759 & 0.0603 & 0.0910 & 0.0828 & 0.0927 & \textbf{0.1029} & $3e^{-9}$ & 11\% \\
 &  & R@40 & 0.1424 & 0.1505 & 0.1482 & 0.1530 & 0.1251 & 0.1778 & 0.1593 & 0.1708 & \textbf{0.1777} & $2e^{-6}$ & 4\% \\
\multirow{-4}{*}{Douban} & \multirow{-4}{*}{$1.4e^{-3}$} & N@40 & 0.0845 & 0.0897 & 0.0887 & 0.0923 & 0.0735 & 0.1090 & 0.0988 & 0.1077 & \textbf{0.1182} & $6e^{-9}$ & 10\% \\

\midrule

&  & R@20 & 0.0103 & 0.0180 & 0.0209 & 0.0225 & 0.0222 & 0.0254 & 0.0314 & 0.0528 & \textbf{0.0578} & $1e^{-11}$ & 9\% \\
&  & N@20 & 0.0072 & 0.0123 & 0.0141 & 0.0154 & 0.0152 & 0.0177 & 0.0213 & \textbf{0.0361} & 0.0290 & $1e^{-13}$ & -20\% \\
&  & R@40 & 0.0170 & 0.0310 & 0.0356 & 0.0378 & 0.0367 & 0.0424 & 0.0519 & \textbf{0.0852} & 0.0752 & $5e^{-11}$ & -12\% \\
\multirow{-4}{*}{Tmall} & \multirow{-4}{*}{$1.2e^{-3}$} & N@40 & 0.0095 & 0.0168 & 0.0196 & 0.0208 & 0.0203 & 0.0236 & 0.0284 & \textbf{0.0473} & 0.0326 & $5e^{-15}$ & -31\%\\

\midrule

\multicolumn{14}{l}{\textit{Sparser datasets}} \\

\midrule

&  & R@20 & 0.0607 & 0.0734 & 0.0782 & 0.0797 & 0.0539 & 0.0834 & 0.0813 & 0.0896 & \textbf{0.0976} & $1e^{-8}$ & 9\% \\
&  & N@20 & 0.0249 & 0.0290 & 0.0315 & 0.0326 & 0.0212 & 0.0334 & 0.0339 & 0.0369 & \textbf{0.0440} & $1e^{-11}$ & 19\% \\
&  & R@40 & 0.0898 & 0.1078 & 0.1155 & 0.1161 & 0.0836 & 0.1196 & 0.1178 & 0.1286 & \textbf{0.1322} & $1e^{-4}$ & 3\% \\
\multirow{-4}{*}{Amazon-beauty} & \multirow{-4}{*}{$7.3e^{-4}$} & N@40 & 0.0308 & 0.0360 & 0.0391 & 0.0400 & 0.0272 & 0.0408 & 0.0413 & 0.0447 & \textbf{0.0511} & $1e^{-11}$ & 14\% \\

\midrule

&  & R@20 & 0.0196 & 0.0552 & 0.0951 & 0.0985 & 0.0869 & 0.0999 & 0.1070 & 0.1578 & \textbf{0.1706} & $2e^{-10}$ & 8\% \\
&  & N@20 & 0.0105 & 0.0298 & 0.0535 & 0.0593 & 0.0528 & 0.0602 & 0.0644 & 0.0935 & \textbf{0.1001} & $8e^{-11}$ & 7\% \\
&  & R@40 & 0.0346 & 0.0810 & 0.1392 & 0.1431 & 0.1276 & 0.1472 & 0.1535 & 0.2245 & \textbf{0.2400} & $5e^{-11}$ & 7\% \\
\multirow{-4}{*}{Gowalla} & \multirow{-4}{*}{$4.0e^{-4}$} & N@40 & 0.0145 & 0.0367 & 0.0684 & 0.0710 & 0.0637 & 0.0725 & 0.0767 & 0.1108 & \textbf{0.1181} & $1e^{-10}$ & 7\%\\

\midrule

&  & R@20 & 0.0328 & 0.0395 & 0.0543 & 0.0542 & 0.0174 & 0.0521 & 0.0501 & 0.0518 & \textbf{0.0588} & $1e^{-10}$ & 14\% \\
&  & N@20 & 0.0169 & 0.0196 & 0.0290 & 0.0288 & 0.0084 & 0.0282 & 0.0270 & 0.0300 & \textbf{0.0368} & $2e^{-9}$ & 23\% \\
&  & R@40 & 0.0439 & 0.0552 & 0.0717 & 0.0708 & 0.0274 & 0.0685 & 0.0655 & 0.0653 & \textbf{0.0706} & $7e^{-8}$ & 8\% \\
\multirow{-4}{*}{Sparser-Tmall} & \multirow{-4}{*}{$8.6e^{-5}$} & N@40 & 0.0195 & 0.0233 & 0.0330 & 0.0327 & 0.0108 & 0.0320 & 0.0305 & 0.0332 & \textbf{0.0395} & $3e^{-9}$ & 19\% \\

        \bottomrule
    \end{tabular}
\end{table*}

\begin{table}[t]
    \centering
    \caption{Dataset statistics.}
    \small
    \label{tab:dataset}
    \begin{tabular}{ccccc}
        \toprule
        Dataset & User \# & Item \# & Interaction \# & Density\\
        \midrule
        Yelp & 29,601 & 24734 & 1,517,326 & $2.1 \times 10^{-3}$\\
        Douban & 13,024 & 22,347 & 397,144 & $1.4 \times 10^{-3}$\\
        Tmall & 47,939 & 41,390 & 2,357,450 & $1.2 \times 10^{-3}$\\
        Amazon-beauty & 22,363 & 12,101 & 198,502 & $7.3 \times 10^{-4}$\\
        Gowalla & 50,821 & 57,440 & 1,172,425 & $4.0 \times 10^{-4}$\\
        Sparser-Tmall & 69,841 & 41,351 & 248,983 & $8.6 \times 10^{-5}$\\
        \bottomrule
    \end{tabular}
    \vspace{-0.15in}
\end{table}
\noindent

\subsection{The Relations between GTE and the Graph and Node-Level Expressiveness Metrics}

As shown in Theorem 1 in Section \ref{sec:graphlevel}, Xu et al. \cite{xu2018powerful} proved that a GNN can have WL-equivalent power of expressiveness on the graph isomorphism test if the aggregation function $h_v^{(k)} = g(\{h_u^{(k-1)}: u \in \mathcal{N}(v)\cup \{v\}\})$ is injective. They further proved that the sum aggregator allows injective functions over multisets. We present an adapted version of the latter conclusion below.

\medskip
\textbf{Theorem 9} (from Xu et al. \cite{xu2018powerful}). \textit{Assume the feature space $\mathcal{X}$ is countable. If the initial features are one-hot encodings, then there exists some function $\phi$ such that the aggregator $h_v^{(k)} = g(\{h_u^{(k-1)}: u \in \mathcal{N}(v)\cup \{v\}\}) = \phi \left( h_v^{(k-1)} + \sum_{u \in \mathcal{N}(v)}h_u^{(k-1)} \right)$ is injective.}

\medskip
In Section \ref{sec:diebipartite}, we demonstrate the effectiveness of a GNN in distinguishing various automorphic nodes on a connected user-item bipartite graph. This is achieved by proving Theorem 4, which shows that when residual connections are implemented and the aggregation function $h_v^{(k)} = g({h_u^{(k-1)}: u \in \mathcal{N}(v)\cup {v}})$ is injective


In GTE, residual connections are employed. Consequently, based on the aforementioned conclusions, if the aggregation function is injective, GTE will exhibit equivalent expressiveness to the Weisfeiler-Lehman (WL) algorithm on the graph-level metric and achieve optimal expressiveness on the node-level metric. It is worth noting that in our implementation, the aggregation function of GTE is not injective, as we simply utilize the identity function as $\phi$. However, according to Theorem 9, it is possible to select $\phi$ carefully, thereby making the aggregation function injective and enhancing the theoretical capability of GTE on both graph and node-level metrics. It is important to acknowledge that while a suitable choice of $\phi$ can render the aggregation function injective, it may also compromise the optimality of GTE on the topological closeness metric. Addressing this issue will be a topic for future research endeavors.

\section{Experiments}
\label{sec:exp}

To evaluate the explainability of topological closeness in recommendation, we perform experiments on real-world datasets to compare GTE, which is optimal on the metric, against various baselines.

\subsection{Experimental Settings}

\subsubsection{\bf Datasets}

\textbf{Yelp}: a widely used dataset containing users' ratings on venues from the Yelp platform. \textbf{Douban}: consisting of book reviews and ratings collected by Douban. \textbf{Tmall}: an e-commerce dataset recording online purchase history of users on Tmall. \textbf{Gowalla}: collected by Gowalla with users' check-in records in different locations. \textbf{Amazon-beauty}: a dataset containing Amazon users' reviews and ratings of beauty products. \textbf{Sparser-Tmall}: a sparser version of the \textit{Tmall} dataset by sampling users and items with less interactions. The statistics are summarized in Table \ref{tab:dataset}.

\subsubsection{\bf Evaluation Protocols and Metrics}

The all-rank protocol is commonly adopted to mitigate sampling bias~\cite{he2020lightgcn,xia2022hypergraph}. For the evaluation metrics, we adopt the widely used Recall@N and Normalized Discounted Cumulative Gain (NDCG)@N, where N = $\{20, 40\}$.\cite{wang2019neural,he2020lightgcn,xia2022self,cai2023lightgcl}. The p-values are calculated with T test.

\subsubsection{\bf Baseline Models}

We adopt the following 8 baselines. We tune the hyperparameters of all the baselines with the ranges suggested in the original papers, except that the size of learnable embeddings is fixed as 32 for all the baselines to ensure fairness.

\noindent

\begin{itemize}[leftmargin=*]

\item \textbf{BiasMF} \cite{koren2009matrix}. This model employs matrix factorization to learn latent embeddings for users and items.




\item \textbf{NGCF} \cite{wang2019neural}. It aggregates feature embeddings with high-order connection information over the user-item interaction graph.

\item \textbf{GCCF} \cite{chen2020revisiting}. It adopts an improved GNN model that incorporates a residual structure and eliminates non-linear transformations.

\item \textbf{LightGCN} \cite{he2020lightgcn}. It implements an effective GNN structure without embedding transformations and non-linear projections.




\item \textbf{SimGRACE} \cite{xia2022simgrace}. It creates an augmented view by introducing random perturbations to GNN parameters.

\item \textbf{SAIL} \cite{yu2022sail}. It adopts a self-augmented approach that maximizes the alignment between learned features and initial features.

\item \textbf{HCCF} \cite{xia2022hypergraph}. It contrasts global information encoded with a hypergraph against local signals propagated with GNN.

\item \textbf{LightGCL} \cite{cai2023lightgcl}. It guides the contrastive view augmentation with singular value decomposition and reconstruction.

\end{itemize}

\vspace{-0.1in}
\subsection{Performance Comparison and Analysis}

The performance results are presented in Table \ref{tab:result}. The number of GNN layers for GTE is set as 3 for all datasets. As shown in the table, GTE can achieve comparable performance to the SOTA GCL models. The mechanism of GTE revolves around effectively discriminating different items based on their topological closeness to a user. Therefore, these results indicate that the proposed topological closeness metric accurately reflects a model's proficiency in performing the recommendation task.


\begin{table}[t]
\setlength{\tabcolsep}{3pt}
    \centering
    \caption{Efficiency comparison in terms of total running time between GTE (w/o GPU) and LightGCL (w. GPU), in minutes.}
    \small
    \label{tab:time}
    \vspace{-0.05in}
    \begin{tabular}{ccccccc}
        \toprule
          & Yelp & Douban & Tmall & Gowalla & Amazon & s-Tmall\\
        \midrule
        LightGCL & 216.12 & 38.54 & 65.03 & 284.12 & 21.06 & 30.24 \\
        \underline{GTE} & 1.85 & 0.64 & 0.84 & 5.11 & 0.26 & 2.10 \\
        \bottomrule
    \end{tabular}
    \vspace{-0.2in}
\end{table}


Furthermore, GTE exhibits superior performance on sparse data. Specifically, while GTE consistently outperforms other baselines, it occasionally performs worse than LightGCL on the three denser datasets. However, on the three sparser datasets, GTE consistently outperforms LightGCL. This discrepancy can be attributed to the reliance of learning-based methods on the availability of supervision signals. In scenarios where there is an abundance of user-item collaborative signals, learning-based methods can more effectively capture complex patterns in the data compared to learning-less or heuristic algorithms like GTE. However, in cases where the interaction history is limited, which often occurs in recommendation systems, the performance of learning-based methods deteriorates more rapidly than that of fixed algorithms.

By capturing the topological closeness between nodes with a learning-less structure, GTE achieves satisfactory performance rapidly and reliably. Since GTE is not subject to any random factors, the p-values in Table \ref{tab:result} are of negligible magnitudes. Table \ref{tab:time} presents an intuitive comparison of the running time of GTE and the efficient GCL baseline LightGCL, where GTE is run on CPU only, and LightGCL is run with an NVIDIA GeForce RTX 3090 GPU.


\vspace{-0.1in}
\subsection{Topological Closeness Aligns with Real Data} \label{sec:topoGNNexp}

\begin{figure}[h]
        \centering
        \vspace{-0.15in}
            \includegraphics[width=\linewidth]{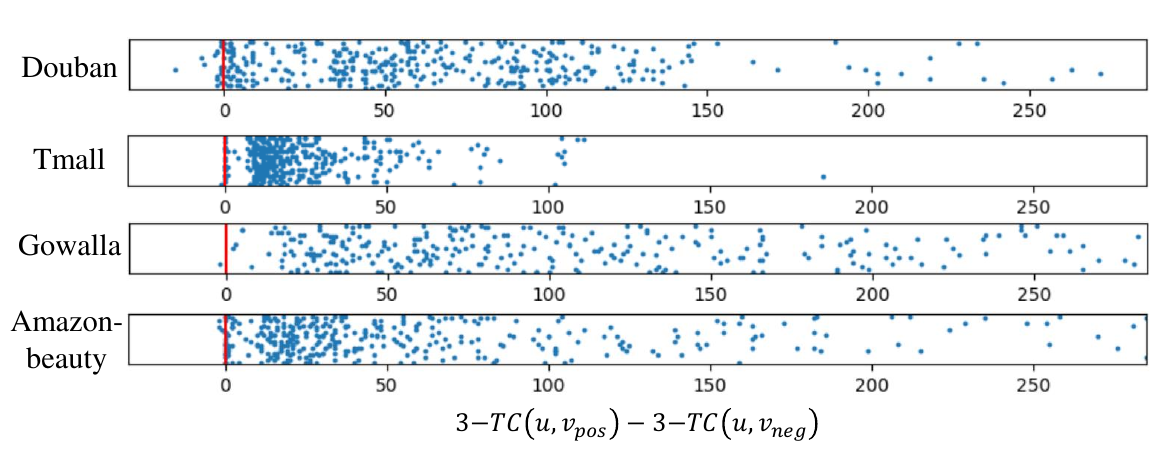}
            \vspace{-0.2in}
            \caption{Difference between the topological closeness of users to their positive items and negative items.}
            \vspace{-0.15in}
            \label{fig:tcdiff}
\end{figure}

To demonstrate the discriminative ability of the topological closeness metric in real datasets, we randomly sample 400 pairs of positive and negative items and calculate the difference between their topological closeness to the target users. The results are plotted in Figure \ref{fig:tcdiff}. It is clear from the plot that the majority of the differences are larger than 0, with only negligible exceptions. This indicates that the positive items have a higher topological closeness to the user than the negative items.

\subsection{Popular GNN Recommenders Perform Poorly on Topological Closeness}

\begin{figure}[h]
        \centering
            \includegraphics[width=\linewidth]{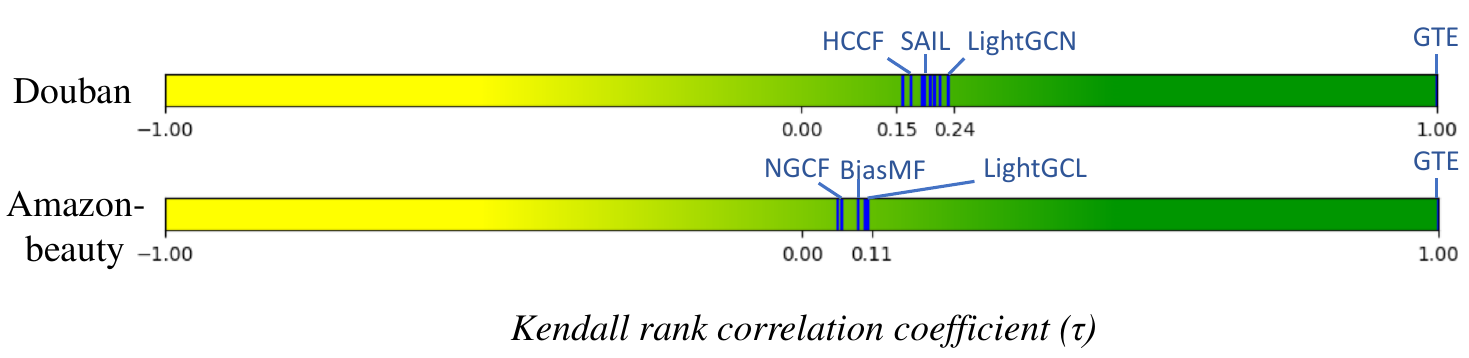}
            \vspace{-0.15in}
            \caption{The ordinal difference (measured in Kendall's $\tau$ coefficient) between the predictions made by the baselines and the ideal predictions based on topological closeness.}
            \vspace{-0.1in}
            \label{fig:tcken}
\end{figure}

Section \ref{sec:theoGNNtopo} provides a mathematical demonstration showing that popular GNN recommender models are unable to fully capture the topological closeness between nodes. Specifically, we sort the items based on the prediction scores generated by GTE and the baseline models. We then calculate the Kendall rank correlation coefficient ($\tau$), a widely used indicator for measuring the ordinal difference between two rankings \cite{kendall1938new}, between these rankings. Since GTE is optimized for topological closeness, its ranking represents the ideal prediction based on this metric. As depicted in Figure \ref{fig:tcken}, the $\tau$ values for the baselines cluster around 0.05 to 0.25. This indicates that the predictions made by the baselines exhibit positive correlations with the ideal ranking, but are significantly misaligned with it.

\section{Conclusion}
\label{sec:conclusion}

In this paper, we provide a comprehensive analysis of GNNs' expressiveness in recommendation, using a three-level theoretical framework consisting of graph/node/link-level metrics. We prove that with distinct initial embeddings, injective aggregation function and residual connections, GNN-based recommenders can achieve optimality on the node-level metric on bipartite graphs. We introduce the topological closeness metric that aligns with recommendation, and propose the learning-less GTE algorithm provably optimal on this link-level metric. By comparing GTE with baseline models, we show the effectiveness of the proposed metric. The presented theoretical framework aims to inspire the design of powerful GNN-based recommender systems. While achieving optimality on all three metrics is challenging, the framework serves as a guiding direction for selecting well-justified model structures.


\clearpage
\balance
\bibliographystyle{ACM-Reference-Format}
\bibliography{sample-base}



\end{document}